\documentclass[12pt,a4papper]{article}
\usepackage{amsfonts,amsmath,amssymb}
\usepackage{mathrsfs}
\usepackage{amscd}

\newcommand{\beq}{\begin{equation}}
\newcommand{\eeq}{\end{equation}}

\date{}
\begin{document}
\title{General structure of the solutions of the Hamiltonian
constraints of gravity}
\author{J. Mart\'in\thanks{Departamento de F\'{\i}sica
Fundamental, Universidad de Salamanca, 37008 Salamanca, Spain,
e-mail: chmm@usal.es}\and Antonio F. Ra\~{n}ada\thanks{Facultad de
F\'{\i}sica, Universidad Complutense, 28040 Madrid, Spain, e-mail:
afr@fis.ucm.es}\and  A. Tiemblo\thanks{Instituto de Matem\'aticas y
F\'{\i}sica Fundamental, CSIC, Serrano 113b, 28006 Madrid, Spain,
e-mail: tiemblo@imaff.cfmac.csic.es}}

\maketitle

\begin{abstract}
A general framework for the solutions of the constraints of pure
gravity is constructed. It provides with well defined mathematical
criteria to classify their solutions in four classes.  Complete
families of solutions are obtained in some cases. A starting point
for the systematic study of the solutions of Einstein gravity is
suggested.

\end{abstract}

\section{Introduction} This paper is devoted to the
search of a general framework for the solutions of the Hamiltonian
constraints of gravity. The canonical description of General
Relativity is an old problem , which goes on acquiring a renewed and
increasing interest as the starting point for the establishment of a
quantum theory of gravity \cite{Arn60,Arn62,DeW67}, which is
probably one of the most ambitious programs of modern physics. It is
an attempt whose final success seems to be still far away.

The idea to extend the methods of quantum field theories to the case
of gravity (perturbative approach) came up against a number of
difficulties, the most evident one being the non renormalizability,
because of which the number of papers on this topic has diminished
in the last years. This is not the case with the non perturbative
approach which, on the contrary, maintain a considerable appeal. A
direct application of the Gauss--Codazzi theorem, allows to split the
four dimensional spacetime into a 3+1 space and time (ADM formalism
\cite{Lic39,Lic44,Cho48,Cho56}, \cite{DeW00,Mis73}), which provides
us with a very useful tool to understand General Relativity in terms
of our direct experience of space and time, {\it i. e.} three
dimensional slices evolving with a parametric time variable. The
final result is a canonical theory with a  Hamiltonian
consisting in a linear combination of first class constraints. Then the logical starting point is to characterize the manifold where the dynamics is defined. This is, precisely, the object of this paper. From the work of these pioneers, this problem lives a
revival with the introduction of gauge theories of gravity
\cite{Heh76,Gro95,Heh95,Tre00}, which are on the basis of the most
recent proposals by Ashtekar and coworkers. They reach a canonical
formalism of gravity, close to standard gauge theories, where the
fundamental dynamical variables are given by the canonically
conjugate pair $(E_i^a, A_{ia})$. Here $E_i^a$ are the SO(3) densitized triads, and $A_{ia}$ the corresponding gauge connections. In the following, latin letters $a,\,b,\,\ldots$ of the beginning of the alphabet are assigned to the coordinates defined in the three
dimensional slices resulting form a suitable foliation of the
space-time, while $i,\,j,\ldots$ of the middle of the alphabet are
internal SO(3) indices (allways in covariant position) running from 1 to 3, as well as $a,\,b,\,\ldots$.

There have been some attempts to formally solve the constraints
\cite{Cap89,Bar95,Cap91,York71}, thoroughly discussed in a previous
reference \cite{Tie06}, in which the close relationship existent between the Ashtekar structure of the constraints and the usual ADM one has been emphasized. Using a suitable choice of the dynamical degrees of freedom, we get, in this work, a general form of the solutions which allows us to classify them in terms of simple and general mathematical conditions. As we will see, these imply restrictions on the three--dimensional metric.

With the purpose to offer a complete and unified version of our
approach, we include, in a first section, a brief review of
reference \cite{Tie06}, adding besides the analysis of some
important properties which reveal themselves useful to construct a
general framework.

\section{Hamiltonian constraints of General Relativity}
The Hamiltonian constraints of General relativity, as derived in the
literature \cite{Bar95b,Imm97,Sam00}, among others, are the weakly
vanishing expressions
\begin{eqnarray} \nabla _aE_i^a &\approx& 0\,,\label{10}\\
E_i^bF_{iab}&\approx&0\,, \label{20}\\
\epsilon
_{ijk}E^a_jE^b_kF_{iab}&\approx&0\,,\label{30}\end{eqnarray}
describing $SO$(3) Gauss law, vector and scalar constraints,
respectively. For simplicity, we have chosen here the
Barbero--Immirzi parameter $\beta =i$, as long as at the classical
level the Einstein's field equations have the same dynamical
contents for any value of $\beta$. $E^a_i=e e_i^a$ are the
densitized triads,  $F_{iab}=2\partial _{[a}A_{|i|b]} +\epsilon
_{ijk}A_{ja}A_{kb}$ is the $SO$(3) field strength tensor, and
$\nabla _aE^a_i=\partial _a E^a_i+\epsilon _{ijk} A_{ja}E^a_k$ being
the local $S$O(3) covariant derivative. Being $A_{ia}$ complex, additional reality conditions are required.

Paying attention to the connections, which are the fundamental
dynamical variables, we have considered in the previous paper
\cite{Tie06} an approach which offers an alternative writing of the
constraints (\ref{10})--(\ref{30}) closer to the ordinary
geometrical ADM language. To do that, we redefine the $SO$(3)
connections as
$$
A_{ia}=\Gamma_{ia}+i k_{ia}\,,
$$
where $\Gamma _{ia}$ is the part of the connection compatible with
the metric and $k_{ia}$ is its intrinsic part, which plays the role
of the extrinsic curvature of the three dimensional slices.

With these elements and after some algebra (the reader is referred
to \cite{Tie06,Romano93,Ort04,Bla02}  for the details), one gets, in
coordinate language, the three equivalent expressions for the
constraints (\ref{10})--(\ref{30})
\begin{eqnarray} k_{[ab]}&\approx& 0,  \qquad \mbox{ Gauss
law}\label{40}\\
D_a(k_b^a-\delta _b^a\,{\rm Tr}\, k)&\approx &0, \qquad  \mbox{
Vector
constraint}\label{50}\\
R^{(3)}-{\rm Tr} (k^2)+({\rm Tr} \, k)^2&\approx &0,\qquad \mbox{
Scalar constraint}\label{60}\,,
\end{eqnarray}
 where $R^{(3)}$ is the
scalar curvature defined in the three dimensional space and the
covariant derivatives are the ordinary Christoffel ones.

It must be emphasized that (\ref{10})--(\ref{30}) and
(\ref{40})--(\ref{60}) are different versions with the same
dynamical contents. Three gauges [$SO$(3), coordinates and
reparametrizations] can be arbitrarily chosen to get the final
result. The problem simplifies drastically after fixing the
reparametrization gauge by imposing the Dirac's maximal slicing
gauge \cite{Hanson76}  ${\rm Tr}(k)=0$ (other choices are possible
as we are going to discuss in the last section), so that it reduces
to solve 
\begin{eqnarray}
D_ak_b^a &\approx &0\,,\label{70}\\
R^{(3)}-{\rm Tr} (k^2)&\approx&0\,,\label{80}
\end{eqnarray}
 where
$k_{ab}$is a traceless symmetric matrix.

It is very easy to verify that the vector constraint (\ref{70})
transforms covariantly in this gauge with respect to the rescaling
of $g_{ab}$ and $k_b^a$, namely
\begin{eqnarray}
g_{ab}&\rightarrow& \varphi\tilde{g}_{ab},\nonumber \\
k_b^a&\rightarrow &\varphi ^{-3/2}\tilde{k}_b^a \nonumber
\end{eqnarray}
what implies
$$D_ak_b^a=\varphi ^{-3/2}\tilde{D}_a\tilde{k}_b^a.$$
Furthermore, the vector constraint always admits  an identical
solution depending only on the metric tensor. In fact, the
Cotton--York tensor $C_b^a$, defined as
\begin{equation}
 C^{ab}= \eta ^{acd} D_c\left[R_d^b- \frac{1}{4}\,\delta _d^bR\right], \label{90}
\end{equation}
is symmetric, traceless and identically conserved $D_aC_b^a\equiv
0$, as required by the vector constraint (\ref{70}).

Consequently, we can establish the general form of $k_b^a$ as the
sum of two terms \beq  k_b^a= \bar{k}_b^a+\alpha C_b^a, \label{100}
\eeq where $\bar{k}_b^a$ stands for any solution of (\ref{70}) and
$\alpha$ is an arbitrary constant.

This can be easily understood, as far as one looks primarily for
solutions for $k^a_b$ in eq. (\ref{70}), which are functionals of an
arbitrary metric, they must be expressible in terms of the metric
tensor and its derivatives up to third order. The reason is that the
first derivatives of the invariant ${\rm Tr} (k^2)$, implicitly
present in (\ref{70}), leads through the scalar constraint to a
derived scalar curvature $R^{(3)}$, which includes up to second
order derivatives of the metric.

On the other hand, the cancellation of the Cotton--York tensor being
the necessary and sufficient condition for a manifold to be
conformally flat, the presence of $C_b^a$ allows us to distinguish,
from the very beginning, conformally flat metrics as a very
outstanding case, as we will see in the following.

\section{The choice of the dynamical variables}
A non negligible part of the problems arising in gravity theories is
related to the search of a suitable set of dynamical variables. In
this section, we consider a parametrization of $k_b^a$ that reveals
itself specially useful to deal with the analysis of the possible
solutions. For this purpose, we recover the triads formalism to
write down $k_b^a$ as
$$
k_b^a= e_i^ak_{ij}e_{jb}\,,
$$
where $k_{ij}$ is an ordinary traceless, symmetric  matrix. In this way the usual $SO(3)$ symmetry of the triads formalism is recovered.
Being symmetric, it can be diagonalized with the help of an
orthogonal matrix, what enables us to deal with its spectral
representation.
$$
k_{ij}=u_i\rho_1u_j+v_i\rho_2v_j+w_i\rho_3w_j\,,
$$
where $\rho_i$ are the three eigenvalues verifying
$\rho_1+\rho_2+\rho_3=0$ (traceless condition) and $u$, $v$ and $w$
are the corresponding eigenvectors. In this way, the five degrees of
freedom of $k_b^a$ arrange as two scalar eigenvalues plus the three
independent parameters associated with the eigenvectors, which are
isomorphic to an $SO$(3) transformation. The last property will be
important in what follows. With these assumptions, the final form of
$k_b^a$ reads \beq
k_b^a=\hat{e}_1^a\,\rho_1\,\hat{e}_{1\,b}+\hat{e}_2^a\,\rho_2\,\hat{e}_{2\,b}+\hat{e}_3^a\,\rho_3\,\hat{e}_{3\,b},\label{110}\eeq
where $\hat{e}_1^a=e_i^au_i$, $\hat{e}_2^a=e_i^av_i$ and
$\hat{e}_3^a=e_i^aw_i$.

We impose now reality conditions on the eigenvalues, which become
constrained to the real roots of the cubic canonical equation. This
restricts the discriminant to be zero or negative. A very convenient
parametrization for both cases is the following

\begin{eqnarray}&& \rho_1= \lambda \cos \omega,\nonumber \\
&&\rho_2=-\frac{\lambda}{2}(\cos \omega -\sqrt{3}\,\sin \omega )\label{120}\\
&&\rho_3=-\frac{\lambda}{2}(\cos \omega +\sqrt{3}\,\sin \omega
)\nonumber
\end{eqnarray}
where $\sin \omega =0$ and $\sin \omega \neq 0$ correspond to null
and negative  discriminant, respectively. Now, we have the essential
elements to attempt a classification of the possible solutions,
paying mainly attention to the vector constraint. It is worth to
emphasize that as a consequence of such a simple parametrization as
(\ref{120}) we have formally solved from the very beginning the
scalar constraint. In fact, a straightforward calculation shows that
${\rm Tr} (k^2)$, which is independent of $\omega$ has the value
${\rm Tr}(k^2)=3\lambda ^2/2$. Therefore, taking into account the
scalar constraint, we have
\begin{equation} \frac{3}{2}\,\lambda
^2=R^{(3)}\label{122}\,.
\end{equation}

\section{Classifying the solutions}
Since there are many and very different solutions of the
constraints, a way must be found to classify them in a general
scheme. For this purpose, we rewrite (\ref{110}) in a more compact
notation as \beq k_b^a=\rho
_{ij}\hat{e}_i^a\hat{e}_{jb}\,,\label{130}\eeq where $\rho _{ij}$ is
the diagonal matrix of the eigenvalues. The vector constraint
becomes then
$$D_a(k_b^a)=D_a(\rho_{ij}\hat{e}_i^a\hat{e}_{jb})=\hat{e}_i^a\hat{e}_{jb}\partial
_a\rho_{ij}+\rho_{ij}\hat{e}_i^aD_a\hat{e}_{jb}+\rho_{ij}\hat{e}_{jb}D_a\hat{e}
^a_i=0\,,$$ which, after multiplication by $\hat{e}_k^b$ gives \beq
\hat{e}_i^a\partial
_a\rho_{ik}-\rho_{ij}\hat{\gamma}_{ikj}+\rho_{ik}\hat{\gamma}_{jij}=0\,,\label{140}\eeq
where the symbols of anholonomy $\hat{\gamma}_{ijk}$ are given as
$$\hat{\gamma}_{ijk}\equiv -\hat{e}_j^b\hat{e}_k^aD_a\hat{e}_{ib}=\hat{e}_{ib}\hat{e}_k^aD_a\hat{e}_j^b$$
and verify
$$\hat{\gamma}_{ijk}+\hat{\gamma}_{jik}=0,\quad \mbox{ and
}\quad\hat{\gamma}_{kjk}=D_a\hat{e}_j^a\,.$$ Note that the covariant
derivative are here the Christoffel ones acting on the coordinate
indices.

It is convenient for the later discussion to develop eq.
(\ref{140}), which leads to
\begin{eqnarray} \hat{e}_1^a\partial _a\rho
_1+\hat{\gamma}_{212}(\rho_1-\rho_2)+\hat{\gamma}_{313}(\rho_1-\rho_3)=0\,,&&\label{150}\\
\hat{e}_2^a\partial _a\rho
_2+\hat{\gamma}_{121}(\rho_2-\rho_1)+\hat{\gamma}_{323}(\rho_2-\rho_3)=0\,,&&\label{160}\\
\hat{e}_3^a\partial _a\rho
_3+\hat{\gamma}_{131}(\rho_3-\rho_1)+\hat{\gamma}_{232}(\rho_3-\rho_2)=0\,.&&\label{170}\end{eqnarray}
Notice that the triads $\hat{e}_i^a$ are defined in terms of the
matrix elements of $k_b^a$. This implies that  the $S0$(3) gauge has
been fixed in (\ref{150})--(\ref{170}) by imposing
$\hat{e}_1^a,\,\hat{e}_2^a,\,\hat{e}_3^a$ to be the eigenvectors of
$k_b^a$ with respect to the metric tensor.

The value of the dynamical variable $\omega$ in the parametrization
(\ref{120}) provides us with a well defined mathematical criterion
to classify all the real solutions. As will be seen, there are four
different classes.

\begin{itemize}

\item {\bf Class A.} It corresponds to $\sin \omega =0$. This is precisely
the case where the discriminant of the secular equation cancels.
Starting from (\ref{150})-(\ref{170}) a bit of algebra leads to the
following system.
\begin{eqnarray} &&\hat{e}_1^a\partial _a \lambda +{3\over 2}\lambda
(\hat{\gamma}_{212}+\hat{\gamma}_{313})=0\,,\label{180}\\
&&\hat{e}_2^a\partial _a\lambda +3\lambda\hat{\gamma}_{121}=0\,,\label{190}\\
&&\hat{e}_3^a\partial _a\lambda
+3\lambda\hat{\gamma}_{131}=0\,.\label{200}
\end{eqnarray}
We recall that $\sin\omega = 0$ is not the only value leading to the last equations, in fact taking $\sin\omega = \pm\sqrt{3}/2$ one recovers (\ref{180})--(\ref{200}) simply by redefining the eingenvalues. So that a general definition of this class can be formulated by imposing $({\rm Tr}\,k^3)^2 = ({\rm Tr}\,k^2)^3/6\,$.

\item {\bf Class B.} The second class corresponds to $\cos \omega
=0$. In this case $k_b^a$ is a singular matrix ({\it i. e.} $\det
(k)=0$), verifying the equations
\begin{eqnarray}
\lambda
(\hat{\gamma}_{212}-\hat{\gamma}_{313})&=&0\,,\label{210}\\
\hat{e}_2^a\partial _a\lambda+\lambda (\hat{\gamma}_{121}+2\hat
\gamma _{323})&=&0\,,
\label{220}\\
\hat{e}_3^a\partial _a\lambda +\lambda (\hat{\gamma}_{131}+ 2 \hat
\gamma _{232})&=&0\,. \label{230}
\end{eqnarray}
As in the previous case one remains in the same class when $\cos\omega = \pm\sqrt{3}/2\,$, leading to the general definition of this class by means the condition ${\rm Tr}\,k^3 = 0$.

\item {\bf Class C. }The third class is the general case, in which the characteristic
equations
can be written as
\begin{eqnarray} \hat{e}_1^a\partial _a\mu +{\mu \over
2}(3-\delta)\hat{\gamma}_{212}+{\mu \over
2}(3+\delta)\hat{\gamma}_{313}&=&0\,,\label{240}\\
\hat{e}_2^a\partial _a[\mu (1-\delta)] +\mu
(3-\delta)\hat{\gamma}_{121}-2\mu
\delta\hat{\gamma}_{323}&=&0\,,\label{250}\\
\hat{e}_3^a\partial _a[\mu(1+\delta)]
+\mu(3+\delta)\hat{\gamma}_{131}+2\mu \delta
\hat{\gamma}_{232}&=&0\,,\label{260}\end{eqnarray} with $\delta
=\sqrt{3}\,\tan \omega\,$ and $\,\mu =\lambda \cos \omega$.

\item {\bf Class D.} Finally a fourth class occurs when we take directly
$\bar k_b^a=0$ in (\ref{110}). Therefore, all the solutions in this
class depend only on the metric tensor. In this case, the
distinction between spaces that are conformally flat and those which
are not acquires a specially relevant role, which will be analyzed
in the following.

\end{itemize}

\section{The choice of the coordinates}
Once characterized the different classes of solutions, we can use
3-diff invariance to choose coordinates. According to a Gauss
theorem, this can be done by fixing the values of the elements of
the metric tensor to render it diagonal. Although other choices are
clearly possible, some of the known relevant solutions can be
written in diagonal form, which on the other hand greatly simplifies
the calculations. We start, therefore, by writing the metric tensor
in the form \beq
g_{ab}=\left(\begin{array}{ccc}a&0&0\\0&b&0\\0&0&c\end{array}\right).
\label{270} \eeq We introduce now a ``natural"\ triad associated
with this form and constructed with the three vectors $e_{1a}=
(\sqrt{a},0,0)$, $e_{2a}= (0,\sqrt{b},0)$ and $e_{3a}=
(0,0,\sqrt{c})$, which are an orthonormal basis with respect to
$g_{ab}$. Any other basis can differ from this one only by the
application of an orthogonal matrix $O_{ij}\,$, so that we can write
\beq \hat{e}_{ia}=O_{ij}\,e_{ja}\,,\label{280} \eeq which expresses
the relation between the eigenvectors of $k_b^a$ and our ``natural"
triad $e_{ia}$. Equation (\ref{280}) allows thus to parametrize all
the arbitrariness of $k_b^a$ simply in terms of a rotation matrix.
Inserting (\ref{280}) in (\ref{180})-(\ref{260}), it is easy to find
the final form of the vector constraint in each class. In this way,
once a metric is adopted, the presence of $O_{ij}$ describes all the
generality of the theory.

As a first simple approach, let us consider the case
$O_{ij}=\delta _{ij}$. A bit of algebra suffices then to find the
expression of the vector constraint in the four classes. In the
case of the class A, it adopts the very simple integrable form
\begin{eqnarray} \partial
_1\log\left[\lambda(bc)^{3/4}\right]&=&0\,,\label{290}\\
\partial
_2\log\left[\lambda a^{3/2}\right]&=&0\,,\label{300}\\
\partial
_3\log\left[\lambda a^{3/2}\right]&=&0\,.\label{310}  \end{eqnarray}
 As is seen, this implies some
restrictions on the fundamental structure of the matrix elements of
the metric tensor, $\lambda a^{3/2}$ being a function of only the
first coordinate and $\lambda (bc)^{3/4}= f(x_2,x_3)$, $f$ being an
arbitrary function of two variables.

The equations in class B can be expressed also in directly
integrable form, namely \begin{eqnarray} \partial
_1\log\left[\sqrt{c/b}\right]&=&0\,,\label{320}\\
\partial
_2\log\left[\lambda\, c\sqrt{a}\right]&=&0\,,\label{330}\\
\partial
_3\log\left[\lambda\, b\sqrt{a}\right]&=&0\,,\label{340}
\end{eqnarray} so that the quotient $c/b$ is independent of the
first coordinate while $\lambda \,c\sqrt{a}$ and $\lambda
\,b\sqrt{a}$ do not depend on the second and third coordinates,
respectively. It must be stressed that that, in both cases A and B,
these conditions are a condensed manner to express families of
solutions. Notice, besides, that since $\lambda =\sqrt{2/3}\,R^{(3)}
$ in Dirac's gauge, they involve only the elements of the metric
tensor.

The study of the third class C leads us to the more complex system
\begin{eqnarray} \partial _1\log\left[\mu (bc)^{3/4} \right]-\delta\, \partial
_1\log\left[(b/c)^{1/4}\right]&=&0,\label{350}\\
\partial _2\log \left [\mu (1-\delta)a^{3/2}\right]+{\delta \over
1-\delta}\,\partial _2\log\left[a/c\right]&=& 0,\label{360}\\
\partial _3\log \left[\mu (1+\delta)a^{3/2}\right]-{\delta \over
1+\delta}\,\partial _3\log\left[a/b\right]&=& 0.\label{370}
\end{eqnarray}
It is clear that this system is, by no means, so easy to solve as
are the previous ones. Nevertheless, assuming in general that $b\neq
c$, $\delta$ can be obtained from the first equation so that, after
substitution in the other two, two conditions can be deduced which
involve the elements of the metric tensor and their first
derivatives and define families of solutions. We recall that this
results correspond to the most simple choice $O_{ij}=\delta
_{ij}$\,, so that this procedure puts in evidence the richness of
solutions of the problem.

\section{The search for the solutions}
In order to find the solutions, we must handle a problem
parametrized in terms of eight degrees of freedom. Two of them are
the eigenvalues of $k_b^a$ although, as emphasized before, one is
formally found by the condition $R^{(3)}=3\lambda ^2/2$. After
fixing the gauge, we have three more which correspond to the
independent elements of the metric tensor. Finally, the remaining
three are parametrized by the defining elements of an orthogonal
matrix, the Eulerian angles for instance. We will obtain in such a
way the four degrees of freedom of pure gravity by solving the
scalar and the vector constraints.

From the mathematical point of view, different possibilities are
open. Nevertheless, it seems convenient to remain close to our
experience of the world, so that the best approach is probably to
use geometry as a primary input.  We will consider, therefore, in
this section several interesting cases in order to test our
treatment in geometrical language.

No doubt, isotropic spaces are obvious candidates for that.
Moreover, as will be seen later, they allow us to propose an
interesting slight modification of our approach. An isotropic
space is described by a diagonal metric tensor that can be written
in spherical coordinates as
$$g_{11} =1/f(r)\,,\quad g_{22}=r^2\,,\quad g_{33}=r^2\sin
^2\theta\,,$$ $f(r)$ being a function depending only on the radial
coordinate $r$.

It is easy to see that the scalar curvature adopts the form
\begin{equation} R^{(3)}={2\over r^2}\,\partial
_r\{r[f(r)-1]\}\,.\label{380}\end{equation} Thanks to our
parametrization, we can thus write immediately the scalar constraint
as
\begin{equation} {3\over 2}\,\lambda ^2={2\over r^2}\,\partial
_r\{r[f(r)-1]\}\,,\label{390}\end{equation} which expresses a
differential relation between $f$ and $\lambda$. At the same time,
this defines $\lambda$ as a function depending only on $r$, a
property that highly simplifies the calculation, as will be seen. As
 in last section, we may start by choosing the matrix
$O_{ij}=\delta _{ij}$, obtaining in this way simple relations
enabling us to construct solutions of Einstein's equations. For
isotropic spaces, a simple calculation shows that the vector
constraint can be written as the system
\begin{eqnarray} \partial _r(r^3\lambda \cos \omega)=0\,, \quad
(a)&&\nonumber \\
\lambda \left[\partial _\theta \cos \omega-{\sqrt{3}\over \sin
^2\theta}\, \partial _\theta (\sin \omega \sin
^2\theta)\right]=0\,,\quad (b)\label{400}\\
\lambda \, \partial _\varphi(\cos \omega +\sqrt{3}\sin \omega)
=0\,,\quad (c)\nonumber\end{eqnarray}

In the case of $\sin \omega =0$
 (class A), (\ref{400}$a$) reduces to
$$\partial (r^3\lambda )=0\,,\quad \mbox{so that}\quad \lambda
=2k_0/r^3.$$ From this and the scalar constraint, it follows
$$\partial _r\left\{r[f(r) -1]+{k_0^2\over r}\right\}=0\,,$$
what leads to the solution
$$f =1-{2m\over r}-{k_0^2\over r^4}\,.$$
It is straightforward to check that the conditions
(\ref{290})--(\ref{310}) are clearly satisfied.

The class D occurs when $\lambda =0$. Since the Cotton--York tensor
$C_b^a$ cancels for isotropic metrics, only the scalar constraint
remains, and it is
$$R^{(3)}={2\over r^2}\,\partial _r \{r[f(r)-1]\}=0,$$
what gives the Schwarzshild solution
$$f =1-{2m\over r}.$$
This function describes a static situation since $k_b^a=0$ in this
case.

 In the class C, both $\sin \omega $ and $\cos \omega $ are
different from zero, so that we write (\ref{400}$a$) and
(\ref{400}$b$) in the form
$$\partial _r\,\omega =\cot \omega\,\partial _r\,\log(r^3\lambda),\qquad \partial
_\theta\, \omega ={-2\sqrt{3}\cot \theta \over 1+\sqrt{3}\cot \omega
}.$$ The integrability condition $\partial _\theta \partial _r
\omega =\partial _r\partial _\theta  \omega =0$ leads to $\tan
\omega =-2\sqrt{3}$, from which some algebra shows that it appears,
curiously, an obstruction to the integrability as the condition
$\cos \theta =0$  on the coordinates. The same happens in class B.

This obstruction $\cos \theta =0$ in classes B and C requires a
comment. One can verify that this is not a property related with
isotropy, but rather a consequence of the choice of the matrix
$O_{ij}$ as a Kronecker $\delta _{ij}$, what puts in evidence the
relevance of the role of the matrix $O_{ij}$ in the construction of
the different solutions. In fact, there are surely families of
solutions corresponding to other choices of the matrix. To
understand this, one can take $O_{ij}$ as a rotation with Eulerian
angles $(\bar \phi,\, \bar \theta, \, \bar \psi)$ such that $\cos
\bar \theta =0$, $\sin \bar \psi =\cos \bar \psi =1/\sqrt{2}$ and
arbitrary value of the azimuthal angle, i. e,
\begin{equation} O= \left (\begin{array}{lll} \sin \bar \phi &-\cos \bar \phi
&0\\ \cos \bar \phi /\sqrt{2}& \sin \bar \phi /\sqrt{2}&
1/\sqrt{2}\\-\cos \bar \phi /\sqrt{2}&-\sin \bar \phi /\sqrt{2}&
1/\sqrt{2}\end{array}\right).\label{410}\end{equation}

The result, not detailed here for simplicity, is that the Eulerian
angle $\bar \phi$ is independent on the azimuthal coordinate
$\varphi$, what does not pose any restriction on the coordinates.

Isotropic spaces have a property suitable to be used in a slightly
different approach, due to the fact that the Ricci tensor becomes
itself diagonal in these spaces. This suggests to parametrize the
solution of the vector constraint in the alternative form
$$k_b^a=\eta ^{acd}D_c M_{db}, $$
where the tensor $M_{db}$ must be symmetric to account for the
property of $k_b^a$ of being traceless. Moreover, it is easy to
show that $M_{db}$  and the Ricci tensor $R_{ab}$ commute,
otherwise $k_b^a$ would not be divergenceless. Finally, the
symmetry of $k^{ab}$ requires that
$$D_a[M_b^a-\delta _b^a\,{\rm Tr}(M)]=0, $$
as is easy to see, just cancelling its antisymmetric part. From
this, it is very easy to obtain again the arbitrariness related to
the presence of the Cotton--York tensor. In fact, if we take simply
$$M_b^a -\delta _b^a{\rm Tr} (M) =G_b ^a,$$
$G_b^a$ being the Einstein's tensor, it turns out that
$$M_b^a=R_b^a-{1\over 4}\,\delta _b^aR $$
reproduces an identical solution. In this way one can work with a
symmetric matrix $M_{ab}$ that can be simultaneously diagonalized
with the Ricci tensor, a property very useful in some cases.

To end this section, we include a brief example of the inverse
problem, starting from a well known solution and reconstructing from
it the different elements of the method. The example is a stationary
four dimensional metric with axial symmetry and Papapetrou's
structure, given as
\begin{equation}
g_{\alpha\beta}=\left(\begin{array}{cccc}g_{00}&0&0&g_{03}
\\
0&g_{11}=a&0&0\\0&0&g_{22}=b&0
\\
g_{03}&0&0&g_{33}=c\end{array} \right),\,\nonumber
\end{equation}
the components of which are functions depending only on $r$ and
$\theta$. The extrinsic curvature is in this case
$$
k_{ab}={1\over \sqrt{-q_{00}}}\,\Gamma^0_{ab}
=\left(\begin{array}{ccc}0&0&\alpha _1\\0&0&\alpha _2\\\alpha
_1&\alpha _2&0\end{array}\right).
$$ Solving now the eigenvalues
 equation with respect to the
three-dimensional restriction of the metric
$$k_{ab}\hat{e}_i^b=\rho _ig_{ab}\hat{e}_i^b\,,$$
we obtain the eigenvalues
$$\rho _1=0,\qquad \rho _2=-\rho _3={\rho \over
\sqrt{abc}}={\sqrt{b\alpha _1^2+a\alpha _2^2}\over \sqrt{abc}}\,.$$
The corresponding eigenvectors can be easily deduced. They are
\begin{eqnarray} \hat{e}_1^a&=&\left({\alpha _2\over \rho},-{\alpha _1\over
\rho},0\right)\,, \nonumber\\
\hat{e}_2^a&=&\left(\sqrt{{b\over 2a}}\,{\alpha _1\over
\rho},\sqrt{{a\over 2b}}\,{\alpha _2\over \rho},{1\over
\sqrt{2c}}\right),\;\;\hat{e}_3^a=\left(-\sqrt{{b\over 2a}}\,{\alpha
_1\over \rho},-\sqrt{{a\over 2b}}\,{\alpha _2\over \rho},{1\over
\sqrt{2c}}\right)\,. \nonumber \end{eqnarray} The matrix $O_{ij}$
relating them with the ``natural"\ triad
$\tilde{e}_1^a=(1/\sqrt{a},0,0)$, $\tilde{e}_2^a=(0,1/\sqrt{b},0)$,
$\tilde{e}_3^a=(0,0, 1/\sqrt{c})$, is
\begin{equation}
O=\left(\begin{array}{ccc} \dfrac{\sqrt{a}\,\alpha
_2}{\rho} & -\dfrac{\sqrt{b}\,\alpha _1}{\rho} & 0
\\[1.8ex]
\sqrt{\dfrac{b}{2}}\,\dfrac{\alpha _1}{\rho} & \sqrt{\dfrac{a}{
2}}\,\dfrac{\alpha _2}{\rho} & \dfrac{1}{\sqrt{2}}
\\[1.8ex]
-\sqrt{\dfrac{b}{2}}\,\dfrac{\alpha _1}{\rho} &- \sqrt{\dfrac{a}{ 2}} \,\dfrac{\alpha _2}{\rho} & \dfrac{1}{\sqrt{2}} \end{array}\right)\,,\nonumber
\end{equation}
in which one recognizes eq. (\ref{410}) if $\sin \bar
\phi=\sqrt{a}\,\alpha _2/\rho$ and $\cos \bar \phi=\sqrt{b}\,\alpha
_1/\rho$. It is therefore a typical B-class solution.

\section{Gauge dependence and concluding remarks}
The main result of this paper is the establishment of a general
framework for the study of the solutions of Einstein's equations in
the case of pure gravity, which gives a classification based on
simple and well defined mathematical grounds. For the sake of
simplicity, we have presented our results in Dirac's ``maximal
slicing gauge", in which the problem becomes particularly simple.
Nevertheless, as is well known, the symmetries of the gravitational
Hamiltonian are non internal. This implies that the different gauges
maybe more or less adapted to the global geometrical properties,
essentially open or closed spaces. As a consequence, a number of
gauge conditions have been proposed (ADM, Dirac or York gauges for
instance) \cite{Arn62} and \cite{Yor71}.

It is not the aim of this paper to discuss the deep dynamical
meaning of a non internal symmetry, we limit ourselves here to look
for a parametrization of the problem, in terms of triads language,
suitable for the study, from a purely mathematical point of view, of
the characterization of the possible solutions.

At this purpose, we consider the constraints (4), (5) and (6) as a
system of equations providing us the natural way to obtain the
extrinsic curvature, using as input a given triad or metric tensor.
This is the reason to adopt the basis in which $K^a_b$ is diagonal,
in such a way that the problem reduces to the calculation of the
three extrinsic curvature eigenvalues.

If we maintain ${\rm Tr} K=\Delta$ different from zero, the
parametrization (12) reads
\begin{eqnarray} &&\rho_1=\lambda \cos \omega + {1\over
3}\,\Delta =\hat{\rho}_1+{1\over 3}\,\Delta,\nonumber \\
&&\rho_2=-{\lambda \over 2}\,(\cos \omega -\sqrt{3}\sin \omega
)+{1\over 3}\,\Delta =\hat{\rho}_2+{1\over 3}\,\Delta ,
\label{43}\\&&\rho_3=-{\lambda \over 2}\,(\cos \omega +\sqrt{3}\sin
\omega )+{1\over 3}\,\Delta =\hat{\rho}_3+{1\over 3}\Delta
.\nonumber\end{eqnarray} A brief calculation shows that the general
form of the constraints (5) and (6) becomes
\begin{eqnarray}
&&\hat{e}_1^a\partial_a\hat{\rho}_1+\hat{\gamma}_{212}
(\hat{\rho}_1-\hat{\rho}_2)+\hat{\gamma}_{313}(\hat{\rho}_1-\hat{\rho}_3)={2\over
3}\,\hat{e}_1^a\,\partial _a\Delta,\nonumber \\
&&\hat{e}_2^a\partial_a\hat{\rho}_2+\hat{\gamma}_{121}
(\hat{\rho}_2-\hat{\rho}_1)+\hat{\gamma}_{323}(\hat{\rho}_2-\hat{\rho}_3)={2\over
3}\,\hat{e}_2^a\,\partial _a\Delta,\label{44}\\
&&\hat{e}_3^a\partial_a\hat{\rho}_3+\hat{\gamma}_{131}
(\hat{\rho}_3-\hat{\rho}_1)+\hat{\gamma}_{232}(\hat{\rho}_3-\hat{\rho}_2)={2\over
3}\,\hat{e}_3^a\,\partial _a\Delta,\nonumber\end{eqnarray} and
\begin{equation} R^{(3)}-{3\over 2}\,\lambda ^2+{2\over 3}\,\Delta
^2=0. \label{45}\end{equation}

We notice that, in this parametrization, the scalar constraint is
independent on $\omega$, thus it can be used to express for instance
$\lambda$ as a function of the three dimensional scalar curvature
and the trace $\Delta$. In this way, the problem reduces to discuss
the vector constraint (\ref{44}) as defined  by a set of partial
differential equations depending on two unknown functions, $\Delta$
and $\omega$. Therefore, to solve them an additional condition
(gauge) in $\Delta$, $\omega$ or both is needed. We emphasize that,
the eigenvalues being scalar quantities, a condition of this kind is
invariant with respect to general coordinate transformations in the
three-dimensional space.

The structure of the vector constraint strongly suggests to fix the
gauge giving a condition on the trace $\Delta$. In this manner, the
vector constraint becomes an ordinary differential equation for
$\omega$. As a matter of fact, the most common choices occurs
precisely when $\Delta$ is taken as a constant function, here of
course ``constant" means a function independent on the three
dimensional space coordinates, a true constant for instance or a
function depending on the time variable (Dirac or York proposals
\cite{Han76}). In both cases the second term in eq. (\ref{44})
vanishes, so that we recover eqs. (\ref{150})-(\ref{170}) and
therefore our classification holds. It must be emphasized that, the
scalar constraint being independent on $\omega$, it is ``formally"
solved, once the value of $\Delta$ is fixed as long as $\lambda$ is
algebraically given in terms of $\Delta$ and the three dimensional
scalar curvatures.

Therefore, it remains only the vector constraint as the natural
criterion to classify the different solutions of the problem, a
classification that, in our scheme, is simply given in terms of the
possible roots of the cubic secular equation. A criterion on the
other hand that remains valid in any possible gauge. In fact, it
must be emphsized that eq. (\ref{44}) can alternatively be used to
directly characterize the different classes. Not only that but, due
to the fact that A and B cases correspond to fixed values of
$\omega$, the problem is formally solved without any additional
condition. However, Dirac's gauge is an useful frame to illustrate
the richness of solutions as long as it describes, in a natural way,
the relevant role of the Cotton-York tensor in the search for three
dimensional geometrical configurations.

\section*{Acknowledgements}

We acknowledge Dr. R. Tresguerres  for useful discussions and to Dr. B. Coll for interesting suggestions on the intrinsic definition of the different classes. J. Martin is in debt to Junta de Castilla y Le\'on for the financial support.

\end{document}